\documentclass[conference]{IEEEtran}
\IEEEoverridecommandlockouts
\usepackage{cite}
\usepackage{amsmath,amssymb,amsfonts}
\usepackage{algorithmic}
\usepackage{graphicx}
\usepackage{textcomp}
\usepackage{xcolor}
\def\BibTeX{{\rm B\kern-.05em{\sc i\kern-.025em b}\kern-.08em
    T\kern-.1667em\lower.7ex\hbox{E}\kern-.125emX}}

\usepackage{url,color,soul}

\begin{document}

\title{Towards Explainable Test Case Prioritisation\\ with Learning-to-Rank Models\\
\thanks{Funding: Grants PID2020-115832GB-I00 and RED2018-102472-T funded by MICIN/AEI/10.13039/501100011033. Andalusian Regional Government (postdoctoral grand DOC\_00944).}
}
\author{
\IEEEauthorblockN{Aurora Ram\'irez, Mario Berrios, Jos\'e Ra\'ul Romero}
\IEEEauthorblockA{\textit{Dept. of Computer Science and Numerical Analysis} \\
\textit{University of C\'ordoba}\\
C\'ordoba, Spain \\
\{aramirez,i12bercm,jrromero\}@uco.es}
\and

\IEEEauthorblockN{Robert Feldt}
\IEEEauthorblockA{\textit{Softw. Eng. Division, Dept. of Computer Science and Engineering} \\
\textit{Chalmers University of Technology}\\
Goteborg, Sweden \\
robert.feldt@chalmers.se}
}
\maketitle


\begin{abstract}
Test case prioritisation (TCP) is a critical task in regression testing to ensure quality as software evolves. Machine learning has become a common way to achieve it. In particular, learning-to-rank (LTR) algorithms provide an effective method of ordering and prioritising test cases. However, their use poses a challenge in terms of explainability, both globally at the model level and locally for particular results. Here, we present and discuss scenarios that require different explanations and how the particularities of TCP (multiple builds over time, test case and test suite variations, etc.) could influence them. We include a preliminary experiment to analyse the similarity of explanations, showing that they do not only vary depending on test case-specific predictions, but also on the relative ranks.

\end{abstract}



\section{Introduction}\label{sec:intro}

As the number of test cases increases, software organisations need to prioritise which ones to execute given recent changes\slash development and test results. In recent years, machine learning (ML) has become a common way to achieve this~\cite{Pan2022}, i.e. training models that prioritise test cases based on historical data and metrics. 

A common problem with ML has traditionally been that predictive performance trumps other qualities when selecting models and approaches. Thus, the size and shape of the trained model is rarely considered and black-box models such as random forests, gradient boosted machines, or neural networks have been the main tools of choice. The models they produce can be hard to understand. In recent years, it has been recognised that this poses several challenges for practical use and the field of eXplainable AI (XAI) has formed to address these challenges. Some researchers even argue that black-box models should not be used in safety-critical applications such as in medicine, finance, or for legal applications~\cite{rudin2019stop}.

In this short paper we consider to what extent and in what ways XAI could add value also in software testing. We use test case prioritisation (TCP) as the case and argue that practitioners will often need to understand why one test case has achieved a certain ranking and why it differs from other ones. In short, our argument is that for impact in practice\slash industry, TCP techniques need to be extended to also provide explanations. A similar argument likely will hold for other types of software testing in the age of AI and ML, and researchers should consider it, more broadly.


\section{Background}\label{sec:background}

\subsection{ML-based test case prioritisation}\label{subsec:tcp}

TCP is a common task in regression testing, where a set of test cases is repeatedly executed to ensure stable functionality and performance in light of new code modifications~\cite{Yoo2012}. Traditionally, TCP strategies rely on heuristics that take test case properties (previous test outcomes, duration, change frequencies, etc.) or code coverage as inputs to decide which test cases are likely to reveal bugs earlier. Recently, we have studied how the amount and diversity of information for TCP can affect its applicability in practice, especially with the rise of ML approaches~\cite{Ramirez2023}.

ML-based TCP has explored different learning approaches, from supervised to reinforcement learning~\cite{Pan2022}. Since TCP is, in essence, a ranking problem, learning-to-rank (LTR) algorithms represent a natural and powerful option. Bertolino et al. compared LTR and reinforcement learning, showing that two LTR algorithms (MART and LambdaMART) were highly effective~\cite{Bertolino2020}. More recently, Yaraghi et al. have conducted an exhaustive study of ML-based TCP, analysing the impact of training with different feature sets, retraining needs, and the trade-offs between feature collection cost and predictive performance~\cite{Yaraghi2022}.  

\subsection{Explainable artificial intelligence}\label{subsec:xai}

XAI promotes the need to explain the decision process of ML systems, and the reasons behind their predictions, in order to increase transparency and confidence in its results. Explainable methods can be oriented towards understanding how the ML model works (\textit{global explanations}) or why a particular result is produced (\textit{local explanations})~\cite{Guidotti2018}. For global explanations, several ML algorithms support calculating the relevance of each feature. Local methods such as Lime, SHAP or Break Down~\cite{Staniak2018} take a sample and a trained model to explain the contributions of the features to the obtained prediction. Other types of local methods generate contrastive or counterfactual explanations~\cite{Chou2022}. The former focus on highlighting the differences between two samples with different predictions, while the latter describe how feature values could change to reverse a prediction (\textit{``what-if''} analysis).

XAI in software analysis has been explored in recent years~\cite{Dam2018}, especially for software defect prediction models~\cite{Jiarpakdee2021,Jiarpakdee2022}. These studies have shown that explanations are useful for practitioners to understand defect prediction models in terms of feature importance (global), why a particular module was predicted to be defective (local), and why a module was defective or not at different points in time (contrastive). We argue that similar benefits hold also for software testing.


\section{Explainability needs in TCP}\label{sec:proposal}

Formally, we define TCP as finding a ranking $R$ that sorts a set of $n$ test cases ($tc_1$...$tc_n$) according to their relevance $r$ (ability to detect faults). Such relevance can be predicted by a LTR model given features ($f_1$...$f_p$) extracted from the test cases, the system under test (SUT) and testing process records. Since a version of the SUT is tested for each build ($b_1$...$b_m$), the TCP process will be at different times ($t_1$...$t_m$), generating a different ranking for each build ($R_1$...$R_m$). Application of a LTR algorithm will therefore produce a ranking $R_i$ for the build $b_i$ using the data collected in previous builds ($b_1$...$b_{i-1}$) for training. Based on these definitions, and our experience working with the software industry on these problems, below we present some scenarios in which explanations of the rankings produced by a LTR algorithm are desirable.

\subsection{Explaining TCP for one build}\label{subsec:xai-tcp-1}

We start with the simplest scenario: the LTR algorithm has produced a ranking $R_i$ for the next build to be tested. In this ranking, the testers might want to know some aspects of the LTR model (global explanations):

\begin{enumerate}
    \item[1A] \textit{Which features influence the relevance predictions of $R_i$ the most?} Recent TCP studies train LTR models with more than 100 features from different information sources, so knowing the most influential ones could reduce the data acquisition effort and help create insights about the development and testing process. In fact, this type of analysis was included in the TCP study by Yaraghi et al.~\cite{Yaraghi2022}, showing the major impact of test case properties (age and execution time) and development aspects (tester experience and contribution).
\end{enumerate}

Also, testers would be interested in understanding the positions of particular test cases (local explanations):

\begin{enumerate}

    \item[1B] \textit{Why is test case $tc_j$ placed in a certain position in the ranking?} Top-ranked test cases might be more attractive for testers to inspect, as they are recommended for early execution. Generating local explanations for the top-ranked test cases will allow comparing the feature importance for specific predictions against the feature importance for the LTR model, as a whole.

    \item[1C] \textit{Why is test case $tc_j$ ranked above another test case $tc_k$}? To answer this question, a method for comparing local explanations is needed. Since the LTR model ranks test cases according to their predicted relevance, one would expect that the positional distance between test cases in the ranking would be reflected in different explanations. We hypothesise that two test cases $tc_j$ and $tc_{j+1}$ in consecutive positions should imply smaller differences in the feature contributions appearing in their explanations compared to two test cases at the extremes of the ranking.
    
    \item[1D] \textit{Why is one test case $tc_j$ likely to fail and another test case $tc_k$ to pass?} Contrastive explanations could help explain the differences between the test cases for which the LTR predicts different verdicts.
    
    \item[1E] \textit{What properties would need to be different to place a test case $tc_j$ higher in the ranking?} In this case, we could make use of counterfactual explanations to discover which features ---and their values--- might increase the likelihood that a test case would be recommended for execution.
\end{enumerate}

If we also had the outcome of (some subset of) the test cases, several of the above scenarios could also be used to build trust in, or create a realistic expectation on, the quality of the prediction model itself.

\subsection{Explaining TCP across builds}\label{subsec:xai-tcp-2}

Next, we briefly describe other scenarios where the generation of explanations for TCP includes a time perspective:

\begin{enumerate}
    \item[2A] \textit{How does the contribution of features to LTR models vary over time?} If the model is retrained after a number of builds to adapt to variations in time-dependent features (e.g. test case age or verdict history), it would be interesting to analyse how global feature contributions vary.
\end{enumerate}

\begin{figure}[!t]
\centerline{\includegraphics[width=0.49\textwidth]{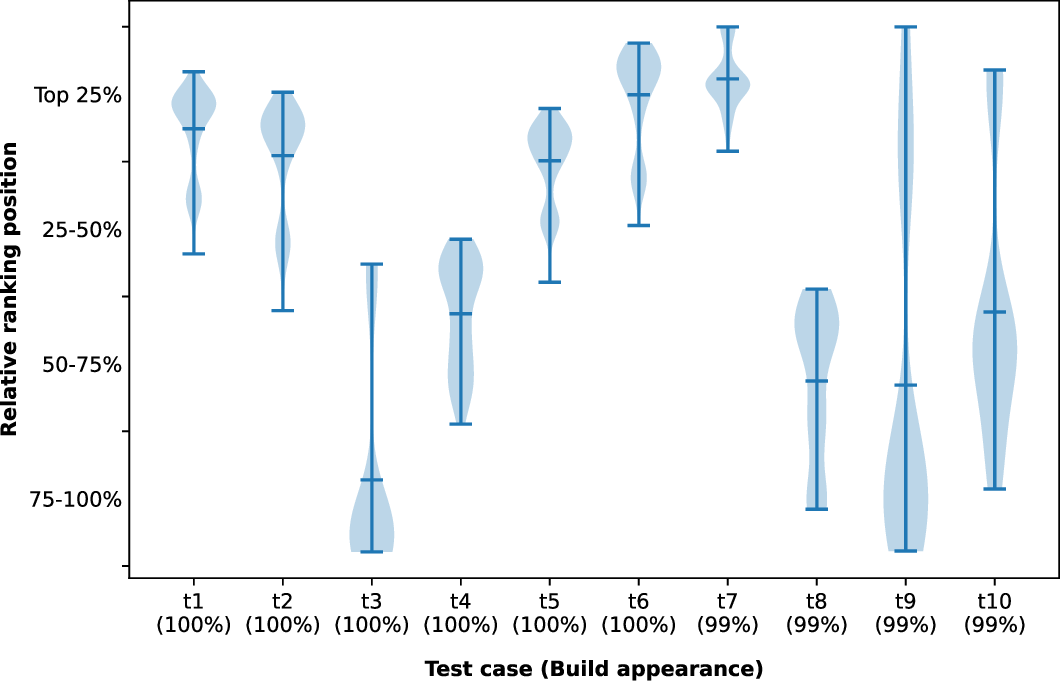}}
\caption{Relative ranking positions of test cases across builds (\textit{angel} system).}
\label{fig:ranking-positions}
\end{figure}

Next, we expose scenarios that need local explanations:

\begin{enumerate}
    \item[2B] \textit{Why did a test case $tc_j$ fail on build $b_{i-1}$ and not on build $b_i$?} Some test cases may fail or pass at different times in the system life cycle depending on the modified parts of the SUT. Based on this assumption, we could compare the explanations generated for different builds to understand this phenomenon.
    
    \item[2C] \textit{Why was a test case $tc_j$ ranked in different positions in the builds $b_{i-1}$ and $b_i$?} The ranking positions are not only influenced by changes in the features used for learning, but also by the number of test cases chosen to test the build, e.g. new test cases are added to the regression test suite.
      
    To illustrate this, Figure~\ref{fig:ranking-positions} shows the relative positions in the ranking of the 10 most frequently executed test cases in the \textit{angel} system, the one selected for our experimentation in Section~\ref{sec:exper}. The ranking positions correspond to the labels in the training sets as proposed by Yaraghi et al.~\cite{Yaraghi2022}. We note that $t_1$ and $t_2$, run on all builds, appear frequently in the top-25 of the rankings. However, $t_9$ and $t_{10}$, also run on almost all builds, appear in very different positions depending on the build. This gives us reason to think that the positions in the ranking are time-dependent, an aspect less explored in XAI, in general, and for LTR algorithms, in particular.
    
    \item[2D] \textit{How do feature contributions vary between different builds for a given test case $tc_j$?} As in scenario 2A, the specific feature contributions that explain the prediction for a test case are expected to change.
\end{enumerate}

\section{Experimentation}\label{sec:exper}

In this section, we take the first steps to experimentally study scenarios 1A, 1B and 1C. Two notebooks to replicate the experiment are available as supplementary material.\footnote{\url{https://github.com/tepia-taxonomy/aist23-workshop}}

\subsection{Methodology}\label{subsec:method}

Here we detail the dataset and the algorithms applied, as well as the process for training and explaining the LTR model.

\subsubsection*{Dataset} We study one of the systems analysed by Yaraghi et al.~\cite{Yaraghi2022}, named \textit{angel},\footnote{\url{https://github.com/Angel-ML/angel}} with 308 builds. It has the highest build failure rate (40\%) after excluding systems with frequent-failing test cases due to configuration issues. The authors collected 150 features, including test case execution records, source code metrics, and coverage information. To obtain test case prioritisations to train with, the test cases were sorted on each failed build based on their verdict and, in case of a tie, execution time in ascending order.

\subsubsection*{Algorithm} LambdaMART~\cite{Wu2017} is the LTR algorithm used for prediction. It combines LambdaRank, a pairwise LTR method frequently used in information retrieval, with the idea of boosted classification trees. We use the implementation available in LightGBM~\cite{Ke2017} with default parameters. We choose this algorithm instead of Random Forest from the RankLib Java library, the one applied by Yaraghi et al.~\cite{Yaraghi2022}, to guarantee language compatibility with explainable methods ---most of them available as Python packages. However, LambdaMART has been applied in other TCP studies showing the second best prioritisation effectiveness and low training time~\cite{Bertolino2020}. Like Yaraghi et al., we set the option to compute global feature contributions (scenario 1A) in the trained model based on their frequency of occurrence in the split nodes.

\subsubsection*{Training and evaluation} We follow a hold-out strategy to train and test the LTR model, using one build for testing, and all previous builds as the training partition. Since LightGBM requires a validation partition, we preserve the last 20\% of builds in the training partition for such a purpose. Model performance is evaluated on the validation partition every 10 iterations in terms of the NDGC (Normalised Discounted Cumulative Gain) metric. This metric does not only evaluate that the highest relevant items (test cases that fail in our experiment) appear at top of the ranking, but also takes item relative relevance in the whole ranking into account. We sort test cases based on the relevance prediction returned by LamdbaMART, using ascending execution time to break ties.

\subsubsection*{Explainable method} After we get the ranking in the test partition, we use Break Down~\cite{Staniak2018} to create local explanations for particular test cases (scenarios 1B and 1C). Break Down assigns a positive or negative contribution to each feature, so that adding all contributions yields the prediction value. The contribution is the mean value of the distribution of predictions obtained for other instances with the same feature value. The goal of this method is to make it easier to understand the explanations by focusing on the contribution of just a few features. For scenario 1C, we will calculate the cosine similarity of the feature contributions for all pairs of test cases in the ranking. We scale the feature contribution values based on the sum of all contributions for each test case prediction, and we keep the contribution signs after adjusting them.

\subsection{Preliminary results}\label{subsec:results}

We run LambdaMART for every failed build of the \textit{angel} system to choose one for our analysis based on its performance (122 builds in total). After sorting all LTR models by NDGC value (in validation) and inspecting the predicted ranking, we select the build $b_{34}$ (35 test cases, 3 of them failed) for testing and the builds $b_1$ to $b_{33}$ for training. The corresponding model has the $5^{th}$ best performance ($NDGC=0.9896$), being the first one able to rank all failed test cases at the top of the ranking. Moreover, the median difference between the true position and the predicted position of all test cases is 1. We briefly summarise the main findings for scenarios 1A-1C based on the selected system and build:

\subsubsection*{Scenario 1A} The five features globally contributing the most to the LTR model are: \textit{OwnersExperience}, \textit{Age} of the test case, \textit{AllCommitersExperience}, \textit{LastExeTime} and \textit{RecentAvgExeTime}. 
Our result is highly aligned with the top-15 of feature contributions reported by Yaraghi et al. considering all systems and builds~\cite{Yaraghi2022}: \textit{Age} ($1^{st}$), \textit{OwnersExperience} ($2^{nd}$), \textit{AllCommittersExperience} ($3^{rd}$), \textit{TotalMaxExeTime} ($4^{th}$), \textit{LastExeTime} ($5^{th}$) and \textit{RecentAvgTime} ($9^{th}$).

\subsubsection*{Scenario 1B} 
\begin{figure}[htbp]
\centerline{\includegraphics[width=0.49\textwidth]{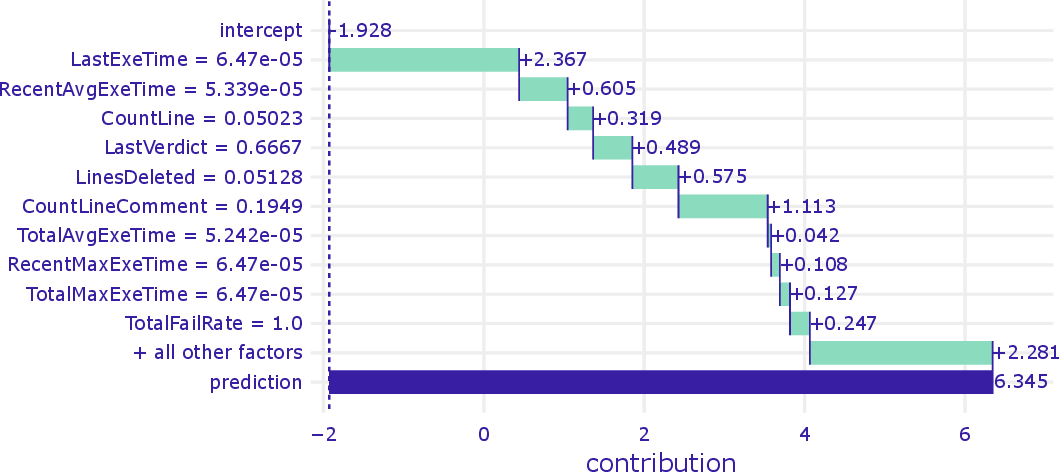}}
\caption{Feature contributions for the test case at top of the predicted ranking.}
\label{fig:local-expl}
\end{figure}

Figure~\ref{fig:local-expl} shows the local explanation generated by Break Down for the test case at the top of the ranking (one of the three failed test cases in the build). Two of the most relevant features (\textit{LastExeTime} and \textit{RecentAvgExeTime}) stood out as important features in the global explanation of the model. Other features with positive contributions are \textit{CountLineComment} ($8^{th}$ in the global feature contribution ranking), \textit{CountLineCodeExe} ($36^{th}$) and \textit{CountLineBlank} ($18^{th}$), which refer to metrics extracted from the test case source code. The fact that important features refer to the test case rather than the entire SUT can help developers to make prioritisation decisions by looking only at test cases.

\subsubsection*{Scenario 1C} Table~\ref{tab:diff-expl} shows the cosine similarity of some pairs of test cases. We focus on the three test cases at the top of the ranking, correctly predicted as failed tests, as well as test cases predicted as non-failed in positions $4^{th}$ (first) and $35^{th}$ (last). We note that the failed test cases have very similar explanations. In fact, the similarity between $tc_1$ and $tc_3$ is the highest among all pairs of test cases. The similarity is reduced when a failed test case is compared to the first non-failed test case ($t_1$ versus $t_4$), and becomes very low for test cases at the extremes of the ranking ($t_1$ versus $t_{35}$). Based on these results, our first approximation seems promising, since similar predictions ---and, by extension, positions in the ranking--- lead to similar explanations. Therefore, developers do not need to inspect the whole ranking of test cases to understand how the LTR model prioritised them.

\begin{table}[htbp]
\caption{similarities of local explanations for pairs of test cases.}
\begin{center}
\begin{tabular}{|c|c|c|}
\hline
\textbf{Test case position}&\textbf{Test case position}&\textbf{Explanation similarity} \\
\hline
$tc_1$ & $tc_2$ & $0.9563$ \\
$tc_1$ & $tc_3$ & $0.9948$ \\
$tc_2$ & $tc_3$ & $0.9509$ \\
$tc_1$ & $tc_4$ & $0.8330$ \\
$tc_1$ & $tc_{35}$ & $-0.7225$\\
\hline
\end{tabular}
\label{tab:diff-expl}
\end{center}
\end{table}

\section{Open issues and conclusion}\label{sec:conclusion}

The following open issues emerge from our study:

\begin{itemize}
    \item Explainability of black-box LTR models have barely been explored, so it is clearly an open line of research. Relevance contributions need different processing than class probabilities, our proposal (scale and similarity) is a first approximation that should be further analysed.
    \item The fact that the LTR model generates a ranking as output also implies that the discrete positions assigned to the test cases can be quite informative in explaining the TCP results. Ideas from explanations in recommender systems~\cite{Zhang2020} could be adapted to the TCP problem.
    \item Ranking positions ---and thus the explanations based on them--- are not directly comparable between builds, since the length of the ranking is different in each build. We need to develop new explainable methods to deal with the evolution of explanations under these conditions.
    \item Interpretable LTR approaches, which provide simpler models and readable results, are also gaining attention~\cite{Zhuang2021}. The advantages of applying interpretable LTR models or black-box LTR models with explanatory capabilities in different software testing situations also demands more research.
\end{itemize}

XAI for software testing, and particularly for TCP, has great potential but it is still in the early stage of research. We seek to contribute to the field with new methods and evaluate them with testers in industrial environments.

\bibliographystyle{IEEEtran}
\bibliography{aist_workshop}

\end{document}